\documentclass[preprint,aps,eqsecnum]{revtex4}
\usepackage{amssymb}
\usepackage{amsfonts}
\usepackage{pstricks}
\usepackage{graphicx}

\begin{document}

\title{Reply to the Comment by Fujikawa and Shrock on the 
Observability of the Neutrino Charge Radius}

\author{J. Bernab\'eu, J. Papavassiliou, and J. Vidal}

\affiliation{
Departamento de F\'{\i}sica Te\'orica and IFIC, \\
Universidad de Valencia-CSIC, \\
E-46100, Burjassot, Valencia, Spain }

\preprint{FTUV-03-0324}

\thispagestyle{empty}

\begin{abstract}
The ``Comment'' presented in hep-ph/0303188 is not even wrong: 
Fujikawa and Shrock miss 
entirely the crucial 
point that the observable quantity associated with the neutrino 
charge radius (NCR) is {\it not} given by the 
(off-shell) one-loop vertex   
$\gamma^{*}\nu \nu$ dictated by the Feynman diagrams in the 
$R_{\xi}$ gauge-fixing scheme.

\end{abstract}
\maketitle

\newpage

It is a well-known fact that in non-Abelian gauge theories 
off-shell Green's functions depend explicitly on the 
gauge-fixing parameter. Therefore, the definition of 
quantities familiar from QED, such as effective charges and 
form-factors, is in general problematic.
Such has been the case with the neutrino electromagnetic 
form-factor and the corresponding NCR. 
The calculational fact 
that the (off-shell) 
one-loop $\gamma^{*}\nu \nu$ vertex (and the NCR obtained 
from it) is 
a gauge-variant quantity has been established beyond any doubt  
in the seventies, in a series of papers cited in \cite{FS} and 
\cite{Bernabeu:2002nw}. 
Based on this observation, it was concluded  
(at the same time and by the same authors)
that ``the NCR'' is not a physical quantity. Of course, 
if something is gauge-dependent it is not physical. But the fact that 
the off-shell vertex is gauge-dependent only means that it just does not 
serve as a reasonable definition of the NCR, it does not mean that 
``an NCR'' cannot be encountered which is physical. Indeed, since then,   
several papers in the literature have attempted to find a 
{\it modified} amplitude, able to give rise to a consistent gauge-invariant 
NCR (see, for example, 
\cite{Lucio:1984mg},\cite{Degrassi:1989ip}, and 
\cite{Papavassiliou:1990zd} just to 
mention some representative cases).
The authors, even though do cite 
some of aforementioned papers, do not
seem to have 
appreciated the 
important ``philosophical'' advancements 
accomplished in these (and later) works, 
which, when properly implemented, allow  
one to evade the aforementioned ``no-go'' claims.

The key issue, at least when seen from the pinch technique 
point of view, is that the gauge-dependent parts of the 
$\gamma^{*}(q)\nu \nu$, 
communicate and eventually cancel against analogous  
contributions concealed inside box-diagrams and self-energy graphs. 
This cancellation mechanism  
allows 
the construction of new amplitudes, which are kinematically
akin to propagators/vertices/boxes, and, unlike the conventional ones, 
are completely independent of the gauge-fixing parameter. 
We emphasize that these
cancellations proceed algebraically, and no integrations over the 
virtual momenta need be carried out. From the ``philosophical'' 
point of view, one should be willing to abandon the convenient,
but at times deceiving, diagrammatic visualization furnished by the 
conventional Feynman graph expansion: photon-related contributions
are not only those that have an off-shell photon explicitly 
sticking out, but those leading to $1/q^2$. 
If one is willing to expand one's notion 
on what a ``photon-like'' contribution is, then one can 
easily evade the old non-observability claims. 
Additional non-trivial steps are of course required in order 
to systematically implement this new point of view; but all such 
steps have been carried out, in excruciating detail, in the 
existing literature, cited in our Physical Review Letter and its 
predecessors.  
With all due respect to the authors, we believe  
that it is important that they read and understand the 
(very detailed and, at times, rather pedagogical) 
literature on this subject.    

What has been accomplished recently in the Physical Review Letter
in question, and the literature cited therein, 
is the proof that there is a 
well-defined, effective Green's function which satisfies gauge-invariance,
renormalization group invariance, and process-independence, and can 
be unambiguously connected to experimentally measurable quantities 
(``Gedanken'' or not). The theoretical methodology allowing this 
physically meaningful definition is that of the pinch technique 
\cite{Cornwall:1982zr}, 
or, equivalently, the background field method in the Feynman gauge. The 
new one-loop three-point 
function $\widehat{\gamma}^{*}\nu \nu$
so constructed, in addition to being independent of the 
gauge-fixing parameter ($\xi$), satisfies a QED-like Ward-identity.   
It is from this latter quantity that the {\it physical} and, as it turns out, 
{\it measurable} NCR will emerge. We insist on the additional 
important point that 
the concept of the NCR is identified with a quantity {\it independent}   
of the particle or source used to probe it; this was proved by 
an explicit construction for unpolarized electrons or for 
right-handed electrons \cite{Bernabeu:2000hf}. 
What depends on the details 
of the target, and seems to be totally unclear to the authors of \cite{FS}, 
is the precise way that the various  
diagrammatic contributions 
conspire in order to always furnish the same unique and gauge-independent 
answer (for example the presence or absence of $WW$ boxes). 
So the logical order should be as follows 
(see also \cite{Papavassiliou:1990zd}): 
(i) Use the pinch technique
to define (at the level of the amplitude) gauge invariant 
self-energies/vertices/boxes. The amplitude which couples electromagnetically
to the target, and does not depend on its details, is to be identified 
with the NCR form-factor 
(ii) Devise an appropriate set of observables,
such that the desired gauge-invariant quantity, 
defined at the previous step,
will be projected out. 

Returning to the present Comment, some final considerations are in order:

(i) The terms of type ($m_e^2/M_W^2)f(\xi)$, stemming from the 
conventionally defined $\gamma^{*}\nu \nu$ vertex, which the authors
claim to be the crux of our ``problem'' are of course irrelevant.
First, it has been shown long ago \cite{Papavassiliou:1990zd}
that 
the pinch technique construction, in general, goes through unaltered 
in the presence of non-vanishing fermion masses. 
Second, for the particular case at hand, we have checked by 
explicit calculation that these terms combine with analogous vertex-like
contributions concealed inside the $WW$ boxes, in order to give rise 
to a completely $\xi$-independent $m_e^2/M_W^2$, which is therefore
genuinely suppressed, and cannot be made arbitrarily large.
By the way, all this happens at the level of a single amplitude,
without need to resort to the observables we have proposed in the
Physical Review Letter in question; resorting to these observables 
is crucial only for projecting out particular gauge-invariant sub-amplitudes,
whose definition, however, is much more general. 
In particular, we do not need to form the differences of observables to 
accomplish the gauge-cancellations, because the cancellations 
are accomplished at a much earlier stage (it is not clear from the 
way the Comment is written whether  the authors appreciate this point). 

(ii) The observable that the authors of \cite{FS} propose, when seen from the 
correct point of view, is in fact nothing but our NCR, whether 
one keeps fermion masses or not. For sure, there seems to be 
a $Z$ and a $\gamma$ part, but in fact the corresponding  
$\xi$-dependent parts, proportional to $q^2$ and $q^2-M_Z^2$, 
respectively, they ``talk'' to each other in a very simple way:
after removing 
the corresponding $Z$ or $\gamma$ tree-level propagator, 
they cancel against each other. 
What is left,
is (a) a ``pure''-gauge-independent $\widehat{\gamma}^{*}\nu \nu$, whose 
contribution is, of course, proportional to 
$q^2$, thus giving rise to the effective contact interaction identified 
as the NCR, and (b)  a ``pure''-gauge-independent $\widehat{Z}^{*}\nu \nu$ 
vertex, which,  due to the QED-like Ward identity it satisfies,   
is also proportional to $q^2$ (and not proportional to $q^2-M_Z^2$),
but is multiplied by a $Z$-propagator.
Therefore, in 
the kinematic limit of interest ($q^2 \to 0$) only  
the contribution of the $\widehat{\gamma}^{*}\nu \nu$ vertex survives. 
So, when varying the neutrino flavors, as the authors do, 
one arrives simply at the difference between our NCR for two different
neutrino flavors.

(iii) If one is willing to use a quantity  
in order to search for New Physics, as the authors advocate, better 
choose one which is observable.

\end{document}